\documentclass[%
 reprint,
 amsmath,amssymb,
 aps,
prb,
]{revtex4-1}

\DeclareMathOperator{\tr}{Tr}
\newcommand*\D{\mathop{}\!\mathrm{d}}

\usepackage{color}

\usepackage{mathtools}
\usepackage{braket}
\usepackage{graphicx}
\usepackage{dcolumn}
\usepackage{bm}
\usepackage{hyperref}

\begin{document}

\preprint{APS/123-QED}

\title{Quantum Quench Dynamics in the Transverse Field Ising Model at Non-zero Temperatures}

\author{Nils O. Abeling}
 \email{nils.abeling@theorie.physik.uni-goettingen.de}
\author{Stefan Kehrein}%
 \affiliation{Institut f\"ur Theoretische Physik, Georg-August-Universit\"at
 G\"ottingen, Friedrich-Hund-Platz 1, 37077 G\"ottingen, Germany.}

\date{\today}

\begin{abstract}
The recently discovered \emph{dynamical phase transition} denotes non-analytic behavior in the real
time evolution of quantum systems in the thermodynamic limit and has been shown to occur in different systems at
zero temperature [Heyl \emph{et al.}, Phys. Rev. Lett. \textbf{110}, 135704 (2013)]. 
In this paper 
we extend the analysis to non-zero temperature by studying a generalized form of the Loschmidt echo,
the work distribution function, of a quantum quench in the transverse field Ising model. 
Although the quantitative behavior at non-zero temperatures still displays features
derived from the zero temperature non-analyticities, it is shown that in this model dynamical phase
transitions do not exist if $T>0$. This is a consequence of the system being initialized in a
thermal state. 
Moreover, we elucidate how the
Tasaki-Crooks-Jarzynski relation can be exploited as a symmetry relation for a global quench or to
obtain the change of the equilibrium free energy density.
\end{abstract}

\maketitle


\section{\label{sec:level1}Introduction}
\begin{figure}[b]
    \centering
    \includegraphics[width=0.7\columnwidth]{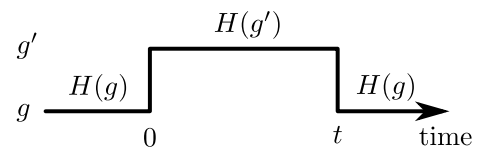}
    \caption{Sketch of a double quench: At $t=0$ the parameter $g$ is quenched to $g'$ such that the
    Hamiltonian $H(g')$ governs the non-trivial time evolution. At $t$ the system is quenched back.}
    \label{fig:sketchdoublequench}
\end{figure}

The past experimental advances in studying closed quantum systems using
cold atomic gases in optical lattices have
spurred the interest and activity in the field of non-equilibrium dynamics.\cite{polkovnikov2011,
kinoshita2006, greiner2002} One benefits from both the
high controllability and time resolution of these systems and the possibility to realize different initial states. 
With this experimental setup at hand one can now address fundamental questions of quantum many-body
physics concerning thermalization and the influence of integrability. The simplest procedure to
drive a system into non-equilibrium is a \emph{quantum quench}, i.\,e., a sudden
change of the Hamiltonian $H(g)$ with only short-range interactions to $H(g')$ where $g$ denotes the 
altered parameter, e.\, g., the
change of an external magnetic field. Having prepared the system with respect to the original Hamiltonian $H(g)$ 
(often in the ground state) the unitary time evolution with the quenched Hamiltonian is non-trivial, since
it starts with a non-thermal superposition of eigenstates. For a more general ramp protocol
$\lambda$ the system parameters set at the specific time $t$ is denoted by $\lambda_t$.\\
To study the quench dynamics one can analyze an experimentally tractable quantity: the performed
work. Unlike equilibrium quantities like the free energy the \emph{inclusive} work $W$ given by the difference of two
consecutive energy measurements, i.\,e. 
\begin{align}
    W = E_m^{\lambda_{\tau}}-E_{n}^{\lambda_0}\!, \nonumber
\end{align}
is well-defined in non-equilibrium.\cite{talkner2007work} Here, $\lambda_{\tau}$ and $\lambda_0$ denote the system
parameters at the respective times of the energy measurements which yielded the eigenvalues $E_n$
and $E_m$ in their instantaneous eigenbasis. 
At non-zero temperature the work depends not only on the entire ramp protocol $\lambda$, but also
on the initial distribution of states and as such is a
stochastic variable with a probability distribution. For this function, which is known under the notion \emph{work
distribution function} (wdf), important theorems describing the fluctuations have been found,
namely, the
famous Jarzynski equality and its generalization, the Tasaki-Crooks relation.\cite{campisi2011} However, due to the
fragility of quantum systems toward the environment and a potential collapse of the wave function, the
theorems have been mostly studied and tested for classical microscopic systems, e.\,g., the
stretched RNA-experiment.\cite{liphardt2002, collin2005, douarche2005} It was only very recently that An \emph{et al.} verified the quantum
Jarzynski equality with a trapped-ion system.\cite{an2015}\\
In this work we calculate the work distribution function for different global ramp protocols and test the 
Tasaki-Crooks relation, thereby
predicting the equilibrium quantity, the change of the free energy density of the corresponding
equilibrium states. \\
Another motivation to study the work distribution function arises when a double quench protocol has 
been implemented (sketched in Fig.\,\ref{fig:sketchdoublequench}). It denotes a quench where the initial and
final Hamiltonian are identical, while the time evolution is governed by $H(g')$, i.\,e.
with a changed parameter $g\to g'$. In this case the Loschmidt echo defined as
$\mathcal{L}(t)=|\mathcal{G}(t)|^2$ is intrinsically contained in 
the work distribution function at $T=0$. Here, the overlap or return probability amplitude
\begin{align}
    \mathcal{G}(t)=\braket{\psi|e^{-iH(g')t}|\psi}\label{eq:lo}
\end{align}
represents the probability amplitude to recover the initial state $\ket{\psi}$ after the quench. In
principle, the initial state 
 can be any state, but is often chosen to be the ground state.\cite{heyl2013}
In other words, the Loschmidt echo measures how much the system is affected by the perturbing quench. 
At the same time, however, the echo represents the probability density that no work is performed on
the system during the double
quench, since the final and
the initial state are identical. The work distribution function can therefore be regarded as the
generalization of the Loschmidt echo, since it also contains the realizations where the system ends
up in a different (excited) state and energy is added or extracted.\cite{silva2008}\\
As a consequence of the aforementioned close relation the work distribution function shares the features
of the Loschmidt echo like the newly discovered \emph{dynamical phase transition} (DPT) which describes non-analytic
behavior in the time evolution (cf.\,\ref{sec:thedpt}).\cite{heyl2013}
This phenomenon was first observed studying the
Loschmidt echo at $T=0$, but also shown to naturally appear in the work distribution function by Heyl
\emph{et al.}. 
Since the initial paper the DPT has been the focus of numerous studies which showed that this
phenomenon also emerges in non-integrable models\cite{karrasch2013, kriel2014} and which treated the
conditions under which they occur.\cite{vajna2014, andraschko2014} Other advances addressed the
question how one can further classify a
DPT and how it arises in the context of DMFT.\cite{canovi2014} Moreover, it was analyzed how local
contributions affect the dynamical free energy and therefore the DPT\cite{fagotti2013}, while other
works revealed a connection of the DPT to the order parameter
dynamics in systems with broken-symmetry phases.\cite{heyl2014}
Due to the close analogy to equilibrium phase transitions a recent development also deals with questions of universality and scaling.\cite{heyl2015universality}
\\
\\
The purpose of this Paper is to investigate the quench dynamics for single and double quench
protocols of the transverse 
field Ising model at non-zero temperature, i.\,e., the effect of the initial state being not the ground state, but 
a thermal state.
Besides the analysis of the corresponding rate functions and the application and verification of the 
Tasaki-Crooks-Jarzynski theorem, we further study whether the DPT persists the change to non-zero
temperatures. 
The procedure starts with the analytic calculation
of the characteristic function of work which can then be transformed to the work distribution
function by applying the G\"artner-Ellis theorem as explained in section \ref{sec:cfow}.
In section \ref{sec:model} we briefly introduce the model, the transverse field Ising model in one dimension, and how it
is solved. We then display the actual calculation of the characteristic function of
work for both the single and the double quench protocol. The results for the rate functions of the
work distribution functions are displayed and discussed in section \ref{sec:results}. Finally, the previous
results are used to verify the Tasaki-Crooks theorem for a global quench for the implemented system
at non-zero temperature. In addition, we demonstrate how the relation can serve 
as a powerful tool to either determine a change in the free energy density or to relate different parts of the
entire work distribution function in section \ref{sec:tasaki}.
\section{\label{sec:cfow}The characteristic function of work}
Since work is a stochastic variable in microscopic systems, it is necessary to consider its
probability density function, the so-called work
distribution function. 
If the principle of microreversibility holds, i.\,e., if
the Hamiltonian is time reversal invariant at any time $t$, the
only remaining ingredient for the formal definition of the work distribution function is the initial
probability distribution of states $p_n^0$ (e.\,g., the canonical distribution).\cite{campisi2011, andrieux2008} 
Having defined the probability to transition from state $n$ into $m$ (also called conditional probability)
$p_{m|n}$, which depends on the ramp protocol
$\lambda$, the work distribution function is given as
the product of the two probabilities and reads
\begin{align}
p[W;\lambda]=\sum\limits_{m,n}\delta\left(W-(E_m^{\lambda_{\tau}}-E_n^{\lambda_0})\right)p_{m|n}[\lambda]p_n^0.\label{eq:wdf}
\end{align}
The edgy brackets emphasize the functional dependence on the ramp protocol $\lambda$.
For complicated ramp protocols the calculation of the wdf can be cumbersome, because the
eigenbasis changes during the quench. While some generic protocols have
been studied,\cite{smacchia2013} the expression simplifies significantly for single or double
quenches. \\ \\
Instead of directly calculating the wdf in Eq.\,\eqref{eq:wdf}, it is often easier to first determine
its generating function, the characteristic function of work $G[u;\lambda]$, defined as the Fourier transform
\begin{align}
    G[u;\lambda]=\int \D W \,e^{iuW}p[W;\lambda]\label{eq:cfowdef}
\end{align}
with $u\in\mathbb{C}$. With $u=iR$, $R\in \mathbb{R}$, the work distribution function then results from a 
saddle-point approximation of the inverse Laplace transformation of $G(R)$. This is achieved by the
G\"artner-Ellis theorem in the context of large deviation theory as explained in section
\ref{sec:results}.\\
The reason to focus on this procedure is the possibility to analytically calculate the
characteristic function of work. It was shown by Talkner \emph{et al.}\cite{talkner2007work} that $G[u;\lambda]$ can be written as a
time-ordered correlation function of exponentiated Hamiltonians as
\begin{align}
    G[u;\lambda]&=\langle
    e^{iuH_{\tau}^H(\lambda_{\tau})}e^{-iuH(\lambda_0)}\rangle\label{eq:cfow}\\
    &=\frac{1}{Z(\lambda_0)}\tr\left(e^{iuH_{\tau}^H(\lambda_{\tau})}e^{-(iu+\beta)H(\lambda_0)}\right).\nonumber
\end{align}
Here, $H_{t}^H(\lambda_t)=U^{\dagger}_{t,0}[\lambda]H(\lambda_t)U_{t,0}[\lambda]$ denotes the
Hamiltonian in the Heisenberg picture at time $t$. The unitary operator $U_{t,0}[\lambda]$ is the usual time
evolution operator which depends on the ramp protocol.\\
For a single
quench the equation above can be simplified further, since $H(t<0)=H(g)$ and $H(t\geq0)=H(g')$
without any time evolution. It follows that 
\begin{align}
    G(u)=\frac{1}{Z(g)}\tr\left(e^{iuH(g')}e^{-(iu+\beta)H(g)}\right).\label{eq:cfowsingle}
\end{align}
At zero temperature the difference of the ground state energies $W_{{\rm min}} = E_0(g')-E_0(g)$ is the minimal work
that can be measured. It is therefore convenient to shift the scale such that $W_{{\rm min}}=0$ and $p(W<0)=0$. 
In order to be able to compare the $T>0$ results to the $T=0$ case we make the same adjustment for
the non-zero temperature calculations. \\
When the ramp protocol is chosen to have a double quench form, Eq.\,\eqref{eq:cfow} yields
\begin{align}
    G(u,t)&=\langle U_{t,0}^{\dagger}(g')e^{iuH(g)}U_{t,0}(g')e^{-iuH(g)}\rangle\label{eq:cfowdouble} \\
    &=\frac{1}{Z(g)}\tr\left(e^{itH(g')}e^{iuH(g)}e^{-itH(g')}e^{-(iu+\beta)H(g)} \right),\nonumber
\end{align}
because the unitary time evolution is now determined by the quenched Hamiltonian.
As the quenches we consider describe \emph{global} quenches, i.\,e. every single
site is affected, the measured work will grow extensively with system size $N$. The quantity that
remains finite is therefore the intensive work density $w=W/N$ which changes the definition of the
characteristic function of work in Eq.\,\eqref{eq:cfowdef} to 
\begin{align}
    G(u)=\int \D w\,e^{iuwN}p(w)=\langle e^{iuwN}\rangle.\label{eq:cfowfinal}
\end{align}
\section{\label{sec:model}The transverse field Ising model}
The model which is studied is the exactly solvable transverse field Ising model in one dimension given by the
Hamiltonian
\begin{align}
    H(g)=-\frac{1}{2}\sum\limits_{l=1}^{N-1}\sigma_l^z\sigma_{l+1}^{z}+\frac{g}{2}\sum\limits_{l=1}^N\sigma_l^x 
\end{align}
with periodic boundary conditions (PBC). It describes a lattice with $N$ spin-1/2 particles that experience
next-neighbor interaction with an overall external magnetic field proportional to $g$. 
The system features a quantum phase transition at the critical
magnetic field $g=g_c=1$. This point separates the quantum paramagnetically ordered phase ($g>1$)
from the regime where ferromagnetic ordering occurs ($g<1$).
To solve the model the Hamiltonian is mapped to a fermionic model via a Jordan-Wigner
transformation\cite{sachdev2011, lieb1961} which leads to a form where the Hamiltonian is split into
different independent momentum sectors such that $H=\sum\limits_{0<k<\pi}H_k$ 
with 
\begin{align}
H_k=\begin{pmatrix}c_k^{\dagger} & c_{-k}\end{pmatrix}\begin{pmatrix}g-\cos k &
        -i\sin k \\ i \sin k & -(g-\cos k)\end{pmatrix}\begin{pmatrix} c_k \\
        c^{\dagger}_{-k}\end{pmatrix}.
\end{align}
The momentum index $k$ is set by the PBC to be $k=\frac{2\pi}{N}m$, $m\in\mathbb{Z}$. 
An additional constant term is ignored, since it does not affect the physics.
The matrix is diagonalized by a fermionic Bogoliubov transformation
\begin{align}
    \begin{pmatrix} c_k \\
            c^{\dagger}_{-k}\end{pmatrix}
        =\begin{pmatrix}
        \cos \theta & i\sin \theta\\
        i\sin \theta & \cos \theta\end{pmatrix}\begin{pmatrix}\eta_k\\ \eta_{-k}^{\dagger}
    \end{pmatrix}
\end{align}
with $\tan 2\theta = \sin k/ (g-\cos k)$ to yield
\begin{align}
    H=\sum\limits_{0<k<\pi}\epsilon_k(g)(\eta^{\dagger}_k\eta_k-\eta_{-k}\eta^{\dagger}_{-k})\label{eq:finalhamiltonian}
\end{align}
with the dispersion relation $\epsilon_k(g)=\sqrt{\left( g-\cos k \right)^2+\sin^2 k}$.\\
From Eq.\,\eqref{eq:finalhamiltonian} one learns that each sub Hilbert space describes a three-level system consisting of the
lowest state without any fermions $\ket{0_k}$ and energy $-\epsilon_k(g)$, a twofold degenerate singly
occupied state $\ket{1_{\pm k}}$ with one fermion with momentum $k$ or $-k$ respectively and zero energy, 
and a fully occupied state $\ket{2_k}$ containing both fermions and
energy $\epsilon_k(g)$. \\
The Bogoliubov transformation can also be used to calculate the quenches.
Depending on the system's parameter the eigenbasis changes over the course of the ramp protocol such that one
needs to translate the new eigenstates in terms of the previous eigenstates. Due to momentum
conservation the singly occupied eigenstates do not change, whereas the unoccupied and fully occupied
states transform via a linear superposition according to
\begin{align}
    \ket{0_k(g)}&=\left(\cos \phi_k +i\sin \phi_k
    \eta_k^{\dagger}(g')\eta_{-k}^{\dagger}(g')\right)\ket{0_k(g')}\label{eq:trafo1}\\
    \ket{2_k(g)}&=\left(i\sin \phi_k +\cos \phi_k
    \eta_k^{\dagger}(g')\eta_{-k}^{\dagger}(g')\right)\ket{0_k(g')}\label{eq:trafo2}
\end{align}
where $\phi_k$ denotes the Bogoliubov angle $\phi_k=\theta(g)-\theta(g')$.
\subsection{Single quench}
The method to calculate the wdf for a single quench uses the characteristic
function of work in Eq.\,\eqref{eq:cfowsingle}. 
Exploiting the simple $k$-sector form of the Hamiltonian one finds for the single quench that
\begin{align}
    G(u)&=\frac{1}{Z(g)}\tr\left(e^{iu\sum\limits_{0<k\pi}H_k(g')}e^{-(iu+\beta)\sum\limits_{0<k<\pi}H_k(g)}\right)\nonumber\\
    &=\frac{1}{Z(g)}\prod\limits_{0<k<\pi}\tr_k\left( e^{iuH_k(g')}e^{-(iu+\beta)H_k(g)}
    \right)\label{eq:res1}
\end{align}
where the partial trace $\tr_k$ only runs over the four eigenstates $\ket{i_k}$ of a $k$-sector. The last step
that rearranges the different $H_k$ becomes possible, because $[H_k(g),H_{k'}(g')]=0$ if $k\neq
k'$. With $Z(g)=\tr e^{-\beta H(g)}=\prod_{0<k<\pi}\tr_k e^{-\beta H_k(g)}$ one obtains 
\begin{align}
    G(u)=\prod\limits_{0<k<\pi}\frac{1}{Z_k(g)}\sum\limits_{i_k}e^{-(iu+\beta)E_{i_k}(g)}\braket{i_k(g)|e^{iuH_k(g')}|i_k(g)}.
\end{align}
\newpage
One way to calculate the explicit form of the equation above is to express the quenched basis states in terms
of the original ones via Eqs.\,\eqref{eq:trafo1} and \eqref{eq:trafo2}. Since the singly occupied
states with zero energy do not change, $\braket{1_{\pm k}(g)|e^{iuH_k(g')}|1_{\pm k}(g)}=1$.
For the remaining two states one finds
\begin{align}
    \braket{0_k(g)|e^{iuH_k(g')}|0_k(g)}=&\cos (u\epsilon_k(g'))\nonumber\\
                                         &-i\cos(2\phi_k)\sin(u\epsilon_k(g'))\\
    \braket{2_k(g)|e^{iuH_k(g')}|2_k(g)}=&\cos (u\epsilon_k(g'))\nonumber\\
                                         &+i\cos(2\phi_k)\sin(u\epsilon_k(g'))
\end{align}
such that eventually 
\begin{align}
    G(u)=\prod\limits_{0<k<\pi}\bigg[\frac{1+\cosh\left( (iu+\beta)\epsilon_k(g)
    \right)\cos\left(u\epsilon_k(g')\right)}{1+\cosh\left(\beta \epsilon_k(g)\right)}\nonumber\\
        -i\,\frac{\sinh\left(
    (iu+\beta)\epsilon_k(g) \right)\cos(2\phi_k)\sin\left(u\epsilon_k(g')\right)}{1+\cosh\left(\beta
    \epsilon_k(g)\right)}\bigg]\label{eq:cfowsinglefinal}
\end{align}
with the partition function for a single mode $Z_k(g)=\sum_{i_k}e^{-\beta E_{i_k}(g)}=2(1+\cosh(\beta \epsilon_k(g))$.
\subsection{Double quench}
The characteristic function for the double quench protocol is calculated in a similar manner, since
a double quench consists of two single quenches and a unitary time evolution in between.
However, this makes it necessary to translate the basis states twice and one finally evaluates
Eq.\,\eqref{eq:cfowdouble} to result in
\begin{widetext}
\begin{align}
    G(u)=\prod\limits_{0<k<\pi}\frac{1+\cosh\left( \beta \epsilon_k(g) \right)\left(
        \cos^2(t\epsilon_k(g'))+\cos^2(2\phi_k)\sin^2(t\epsilon_k(g'))
    \right)+\sin^2(2\phi_k)\sin^2(t\epsilon_k(g'))\cosh\left( (2iu+\beta)\epsilon_k(g)
\right)}{1+\cosh(\beta \epsilon_k(g))}.\label{eq:cfowdoublefinal}
\end{align}
\end{widetext}
\subsection{The dynamical phase transition at $T=0$}
\label{sec:thedpt}
In the previous sections it was shown how the work distribution function is related to the
characteristic function of work for a general complex $u$ and how it can be calculated for the
one-dimensional transverse field Ising model. 
To see how the DPT appears in the work distribution function one first has to understand the
connection to the Loschmidt echo. \\
One way follows a similar path as the derivation by Heyl \emph{et al.} in their original
paper.\cite{heyl2013} 
Motivated by the similarity between the canonical partition function of an equilibrium system 
$Z(\beta)=\tr e^{-\beta H}$ and the Loschmidt overlap they studied the boundary partition function 
\begin{align}
    Z(z)=\braket{\psi_0|e^{-zH}|\psi_0}\label{eq:bpf}
\end{align}
with $z\in\mathbb{C}$ and $\ket{\psi_0}$ denoting the ground state. 
This expression represents the Loschmidt overlap when $z=it$, i.\,e.,
on the imaginary axis.
Due to exponential scaling with system size this quantity takes on a large deviation 
form $Z(z)\sim e^{-Nf(z)}$ in the thermodynamic limit such that one defines the free energy density 
\begin{align}
    f(z)=-\lim\limits_{N\to\infty}\frac{1}{N}\ln Z(z) \label{eq:fed}
\end{align}
where $N$ is the system size.
Analogously to equilibrium phase transitions where
non-analytic behavior of the free energy density results from zeros of the partition function in the
complex temperature plane, Heyl \emph{et al.} showed that a similar phenomenon occurs when analyzing the
dynamics of the Loschmidt overlap. In the
thermodynamic limit the zeros of the boundary partition function, the so-called \emph{Fisher zeros},
coalesce into lines, which cross the time axis under certain conditions.\cite{heyl2013} This leads to non-analytic behavior
in the free energy density, which is just the rate function of the Loschmidt overlap, at
critical times $t^*_n$ with
\begin{align}
    t^*_n=\frac{\pi}{E_{k^*}(g')}\left( n+\frac{1}{2} \right),\;\;\;n=0,1,2,\ldots
\end{align}
where the dispersion relation is given by $E_k(g)$ and $k^*$ is determined by $\cos
k^*=(1+gg')/(g+g')$. The close analogy to equilibrium phase
transitions inspired the authors to call 
this phenomenon \emph{dynamical phase transition}. In the introduction we argued that the Loschmidt echo 
can be identified with the work distribution
function where no work is performed on the system, i.\,e. $p(w=0)$. Therefore the DPT also appears
as non-analytic behavior in the corresponding rate function $r(w,t)$ of the work distribution
function. Mathematically, one can indeed show that the singularities are the same for a double
quench protocol.\\
Another way to recognize the close relation between the Loschmidt echo and the work distribution
function is to analyze the characteristic function of work $G(u)$. If, for example,
one considers a single quench which requires two energy measurements in order to determine the work, then the corresponding characteristic function of work is given by 
Eq.\,\eqref{eq:cfowsingle}. Taking the zero temperature limit $\beta \to \infty$ reduces the trace over all states to
a single term with the ground state, such that one ends up with  
\begin{align}
    G(u)=\langle \psi_0 | e^{iu(H(g')-E_0(g))}|\psi_0\rangle.
\end{align}
It readily follows with $u=t$ that $G(t)=\mathcal{G}(t)^*$ where the energy scale of $H(g')$ has
been shifted by the constant ground state energy $E_0(g)$.\cite{silva2008} This shift is a remnant of the
definition of the work distribution function in Eq.\,\eqref{eq:wdf} where $E_n^{\lambda_0}$ was not set
to 0 for $n=0$. 
\\
As the work distribution function treats all possible measurement outcomes it is more difficult to
relate the general double quench wdf to the Loschmidt echo. From the previous considerations one
assumes a connection to the case where no work is performed ($W=0$) in the zero temperature limit.
The easiest way to establish the link is to start with the general wdf in Eq.\,\eqref{eq:wdf} where
the time evolution is contained within the transition probability $p_{m|n}$ and perform the
zero temperature limit $\beta\to\infty$ ($\lambda_{\tau}=\lambda_0=g$). The limit simplifies the
initial distribution of states $p_n^0=\delta_{n0}$ where only the ground
state ($n=0$) survives. With $p_{m|n}=|\braket{E_m(g)|e^{-iH(g')t}|E_n(g)}|^2$ this finally yields
\begin{align}
    p(W,t)=&\sum\limits_m\delta(W-(E_m(g)-E_0(g)))\nonumber\\
    &\times|\braket{E_m(g)|e^{-iH(g')t}|E_0(g)}|^2.
\end{align}
For $W=0$ the equation further reduces to the Loschmidt echo $\mathcal{L}(t)$ demonstrating the
close relation of the two quantities. Therefore, we expect that any non-analytic behavior in the rate function of
$\mathcal{L}(t)$ can be found by studying the dynamics of the rate function of the work distribution function.
\section{\label{sec:results}The work distribution function}
In the previous part the characteristic function of work was calculated for both the single and double
quench. Now, in the last step, one needs to calculate the wdf. Similar to the boundary partition
function the work distribution function $p(w)$ is expected to show
large deviation behavior, i.\,e. $p(w)\propto e^{-Nr(w)}$, with some non-negative rate function
$r(w)$.\cite{gambassi2012} Besides containing the mean value $\bar{w}$ where $r(\bar{w})=0$ and thus
$p(\bar{w})$ is maximal, the rate function provides additional information about
fluctuations in its tails. The underlying theory, the \emph{large deviation theory}, is therefore
sometimes regarded as an extension of the law of large numbers and the central limit
theorem.\cite{touchette2009} To calculate the rate function of the wdf via the characteristic
function of work one applies the G\"artner-Ellis theorem as explained in the
following.\cite{touchette2009}\\
Setting $u=iR$, $R\in\mathbb{R}$, in Eq.\,\eqref{eq:cfowfinal} one can define the negative scaled cumulant
generating function of $w$ 
\begin{align}
    k(R)=-\lim\limits_{N\to\infty}\frac{1}{N}\ln G(R)
\end{align}
in analogy to the free energy density in Eq.\,\eqref{eq:fed}.
If $k(R)$ exists, $G(R)$ is said to obey a large deviation principle, i.\,e., $G(R)\propto
e^{-Nk(R)}$, with a rate function $k(R)$.\cite{touchette2009} Moreover, it follows that $k(R)$ is
always concave and continuous inside the relevant domain of definition.
The function $k(R)$ simplifies significantly, because $G(R)$ splits into products over $k$. 
Thus,
\begin{align}
    k(R)&=-\lim\limits_{N\to\infty}\ln\prod\limits_{0<k<\pi}G_k(R)\nonumber\\
        &=-\lim\limits_{N\to\infty}\sum\limits_{0<k<\pi}\ln G_k(R)\nonumber\\
        &=-\int_0^{\pi}\frac{\D k}{2\pi}\,\ln G_k(R).\label{eq:nscgf}
\end{align}
Having showed the existence of $k(R)$ the G\"artner-Ellis theorem states that, if $k(R)$ is
differentiable, the probability distribution $p(w)$ also takes on the large deviation form. The
corresponding rate function $r(w)$ is then obtained by a Legendre-Fenchel
transformation\cite{touchette2009} via
\begin{align}
    r(w)=-\inf\limits_{R\in\mathbb{R}}(wR-k(R)).\label{eq:lf}
\end{align}
Here, the infimum is evaluated within the domain of definition of $k(R)$. 
\begin{figure}[tb]
    \centering
    \includegraphics[width=\columnwidth]{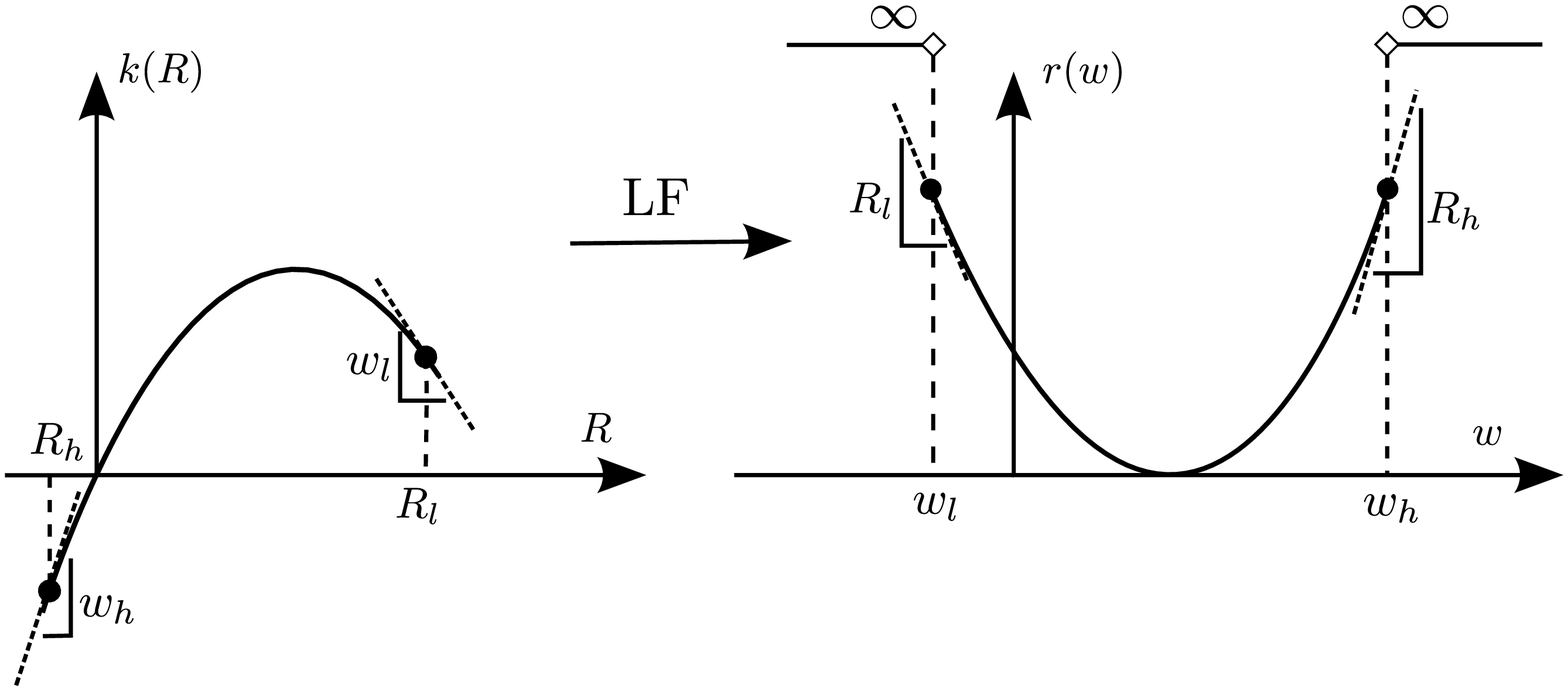}
    \caption{Sketch of the Legendre-Fenchel (LF) transformation for the negative scaled cumulant
        generating function $k(R)$ for a single or double quench in the
        transverse field Ising model. At $R_h$ and $R_l$ $k(R)$ takes on a linear form with slope $w_h$ 
        or $w_l$, respectively.
        These slopes and their positions determine the interval with finite values as depicted on
        the right.}
    \label{fig:lf}
\end{figure}
In Fig.\,\ref{fig:lf} a sketch of the LF transformation is depicted to show how the shape of $k(R)$ determines the 
properties of $r(w)$. Except the precise quantitative behavior one can already deduce several
generic features of $r(w)$ by analyzing $k(R)$. In the region where $k(R)$ is strictly concave,
i.\,e. not even linear, there is a duality between the two related functions that transforms the slope $\tilde{w}$ of $k(R)$ at
position $\tilde{R}$ into the slope $\tilde{R}$ of $r(w)$ at position $\tilde{w}$ and
vice versa.\cite{touchette2009}
It follows that a function $k(R)$, which is strictly concave on the entire axis, would lead to a rate function that
is not confined to some interval, because the positive and negative slopes are
unbounded. While this can be the case for a free bosonic field theory,\cite{gambassi2012} we expect
a different behavior for the transverse field Ising model as we will see shortly. \\
If the negative scaled cumulant generating function $k(R)$ has a linear asymptotic behavior, it is
not strictly concave and the duality does not hold anymore. Let us assume that the slopes of $k(R)$ are bounded 
and have a maximal and minimal value which are already reached for finite $R$ as sketched in Fig. \ref{fig:lf}.
Then, the LF transformation yields for the linear parts a
negative infimum value which is infinite such that $p(w>w_h)=p(w<w_l)=0$. Moreover, the rate
function of the wdf will be finite at the boundaries as the linear behavior is
not only asymptotically reached, but for some finite $R$.
Summing up, the expectation is to observe a rate function which has a similar shape as the sketch in
Fig.\,\ref{fig:lf} on the right.\\ 
Having explained how the shape of the rate function and its corresponding negative scaled cumulant
generating function are mathematically related via the LF transformation we still lack a physical motivation why
the rate function $r(w)$ should look as depicted. 
Indeed, there has to be a minimum and a maximum in the work 
density due to the fermionic properties of the system. Let us consider, for example, a first measurement 
where all $k$-sectors are in the highest state, i.\,e. the doubly occupied state, while a second 
measurement shows no sector is occupied. Then, the difference of the two
measurements will represent the maximal energy that can be extracted from the system and therefore
be the minimal possible work density. More energy cannot be extracted, because there are no
further fermions that could be excited initially. An analogous argument holds for the maximal work
density.\\ 
In the following analysis the rate functions for the single and double quench protocols are studied
in detail. The results are obtained by numerically computing both the Legendre-Fenchel
transformation in Eq.\,\eqref{eq:lf} and its argument, the integral in Eq.\,\eqref{eq:nscgf} over the analytic
expression in Eq.\,\eqref{eq:cfowsinglefinal} or \eqref{eq:cfowdoublefinal}.
The explicit time-dependence that
appears when considering the double quench protocol translates into a time-dependent $k(R,t)$ and
finally $r(w,t)$. 
\subsection{Rate function for the single quench}
\begin{figure}[tb]
    \centering
    \resizebox{\columnwidth}{!}{\includegraphics{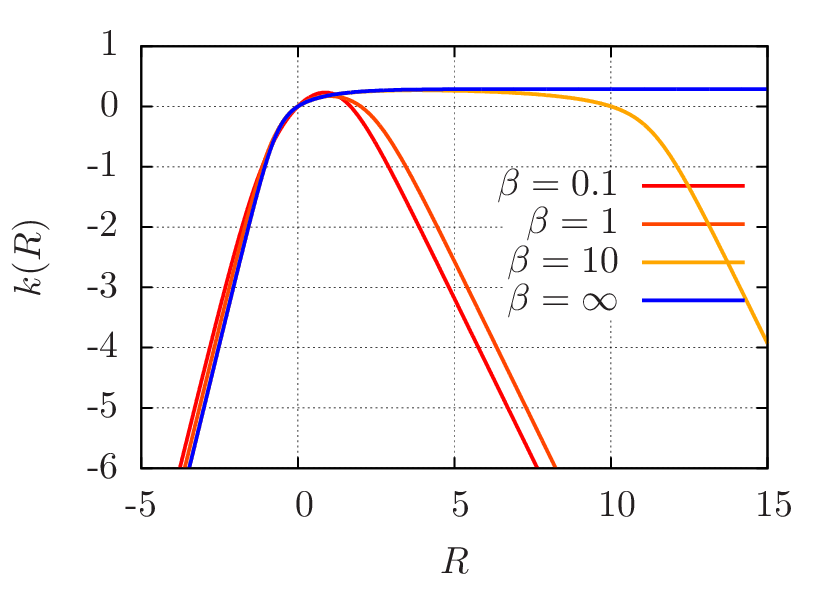}}
    \caption{Negative scaled cumulant generating function $k(R)$ of a single quench from $g=0.5$ to
    $g'=2$ at different inverse temperatures $\beta$. All functions at different non-zero
    temperature share an identical asymptotic behavior for $R\to\infty$ or $R\to-\infty$,
    respectively. 
    It follows that the corresponding rate function is confined to an interval on the work density
    axis as expected. When the temperature is very low, plateaus are created which become more
    pronounced with decreasing temperature until the zero temperature limit (blue) shows an asymptote with vanishing slope for $R\geq\tilde{R}$. This ensures that $p(w<0)=0$.}
    \label{fig:nscgfsingle}
\end{figure}

The first part of the analysis concentrates on studying the single quench protocol. 
At $T=0$ the minimal measurable work density is the difference of the ground
state energy densities of $H(g)$ and $H(g')$ given by $\Delta = \epsilon^0(g')-\epsilon^0(g)$ with
\begin{align}
    \epsilon^0(g)=\frac{E^0(g)}{N}&=-\frac{1}{N}\sum\limits_{0<k<\pi}\epsilon_k(g) \overset{N\to\infty}{=}-\frac{1}{2\pi}\int\limits_0^{\pi}\D
    k\,\epsilon_k(g)\nonumber\\
    &=-\frac{1+g}{\pi}\text{Ell}_2\left( \frac{4g}{(1+g)^2} \right)\label{eq:energylevel}
\end{align}
and Ell$_2(x)$ denoting the elliptic integral of second kind. 
Shifting $w$ by $\Delta$ ensures that $w=0$ corresponds to the smallest accessible
work measurement such that $p(w<0)=0$. For the following analysis of the 
rate function at $T>0$ this energy shift is always included to be able to compare the results.\\
First of all, one has to check for the differentiability of $k(R)$ such that the G\"artner-Ellis theorem
is applicable. Since the argument of the logarithm of the respective integrand in Eq.\,\eqref{eq:nscgf} given by
Eq.\,\eqref{eq:cfowsinglefinal} is an analytic function, this is valid at non-zero temperatures. 
It does not allow for poles,
roots or non-analyticities. \\ 
The negative scaled cumulant generating function $k(R)$ for
different inverse temperatures $\beta$ is depicted in Fig. \ref{fig:nscgfsingle}.
Studying this function we can already deduce its main features using the previous considerations.
Besides the zero temperature case, where the right asymptote
has a slope equal to zero, the asymptotes of $k(R)$ at any other temperature share a common negative slope. 
This behavior is
responsible for the rate function at $T=0$ to be confined to values $w\geq0$. On the
left side, however, all temperatures have the same positive slope for $R\ll0$. 
The analytical method to study the asymptotic behavior is the analysis of Eq.\,\eqref{eq:nscgf} for
$R\to\pm\infty$. 
One finds that for $T>0$
\begin{align}
    k(R)\sim\begin{cases} \,\;\;2\epsilon^0(g)R+C_{\infty} & {\rm for}~R\to\infty\\
        -2\epsilon^0(g')R+C_{-\infty} & {\rm for}~R\to-\infty
    \end{cases}\label{eq:limits}
\end{align}
with constants $C_{\pm\infty}$ and (negative) ground state energy densities $\epsilon^0(g)$ and $\epsilon^0(g')$. 
This directly translates into an identical upper bound for all rate functions $r(w)$ with any finite
temperature and a common negative lower bound for all temperatures $T>0$.
Consequently, it follows that at non-zero temperature
\begin{align}
    p(w)=0\;\;\;{\rm for}~w<2\epsilon^0(g)~{\rm and}~w>-2\epsilon^0(g').\nonumber
\end{align}
In other words, it is possible that negative work can be measured if $T>0$, although being highly 
improbable for very low temperatures. 
Microscopically, this can be realized by a quench where many initially excited $k$-sectors end up in
states with lower energy, thus extracting energy from the system.  
This does not violate the Second Law of
Thermodynamics, since the mean value $\bar{w}$ where $r(\bar{w})=0$ is always larger than 0.
The limits in Eq.\,\eqref{eq:limits} show that the pre-quench parameter $g$ determines the lower
bound, while the post-quench value $g'$ decides upon the upper bound. 
This can be understood as follows: If, for example, the system
begins in the highest excited state (all $k$-sectors are doubly occupied) the energy density is given as
$\epsilon^{\text{h}}(g)=-\epsilon^0(g)$. To measure the minimal
work one needs to extract maximal energy from the system, which can be achieved by 
\begin{figure}[tb]
    \centering
    \resizebox{\columnwidth}{!}{\includegraphics{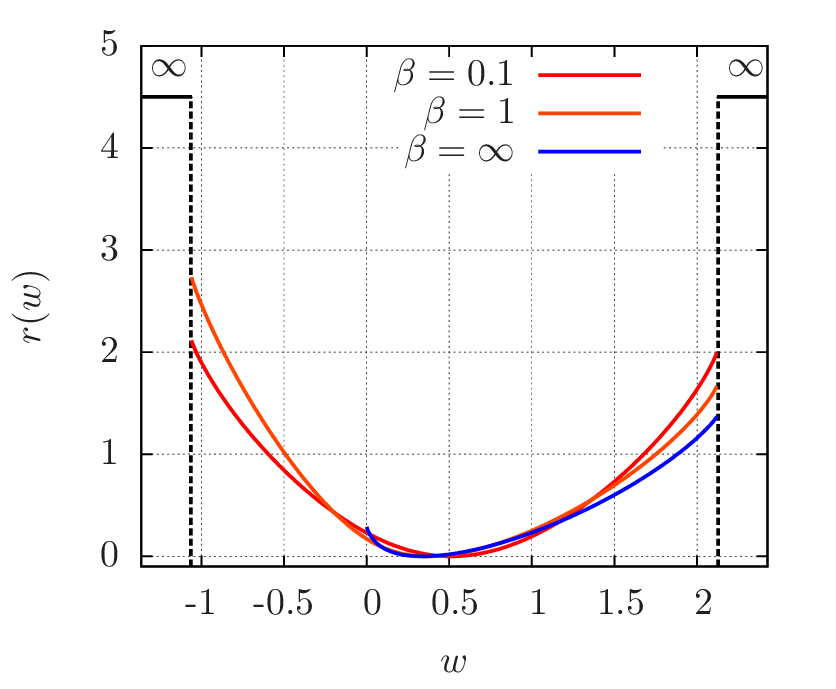}}
    \caption{The rate function $r(w)$ of the work distribution function for a single quench from $g=0.5$ to 
    $g'=2$ at different inverse temperatures $\beta$. For non-zero temperatures the rate function is
    finite within some interval determined by the ground state energy densities of the pre- and post-quench Hamiltonian.
For larger or smaller work densities, the rate function is infinite (black lines) such that the corresponding
work distribution function vanishes. The $T=0$ ($\beta=\infty$, blue line) case shows a peculiarity: The rate
function is already infinite for $w<0$, such that $p(w<0)=0$. This results from the fact that
initially only the ground state of the pre-quench Hamiltonian is occupied.}
    \label{fig:ratefuncsingle}
\end{figure}
\begin{figure}[!h]
    \centering
    \resizebox{\columnwidth}{!}{\includegraphics{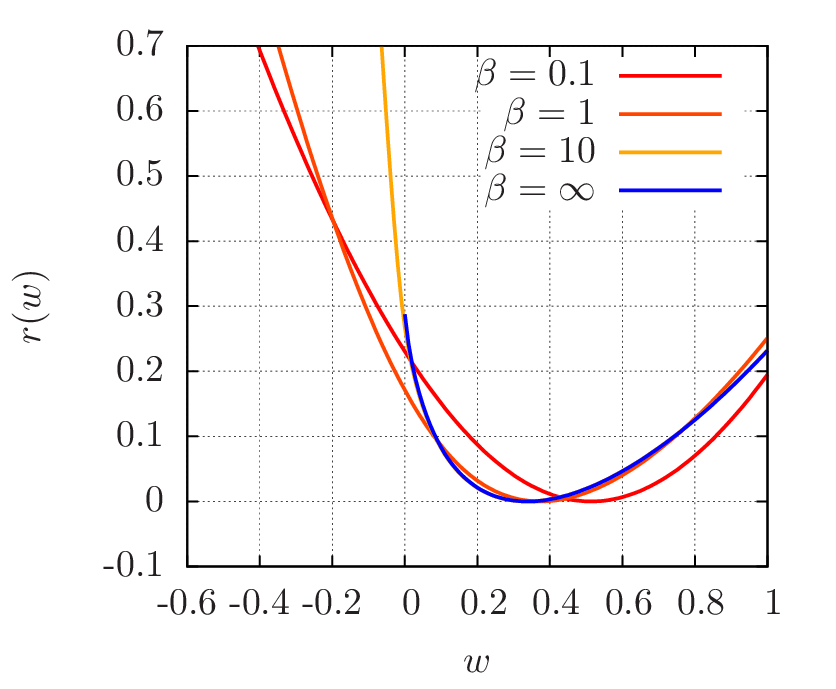}}
    \caption{Zoom into Fig. \ref{fig:ratefuncsingle} showing the rate function $r(w)$ of the work
distribution function for a single quench from $g=0.5$ to $g'=2$ at different inverse temperatures
$\beta$. In addition to the previous data the curve for $\beta=10$ has been added to demonstrate the
steep ascent of the rate function for low temperatures. One notices that the difference between the
low temperature and zero temperature results are already too small to be adequately resolved.}
    \label{fig:ratefuncsinglezoom}
\end{figure}
quenching the system to the ground state of the post-quench Hamiltonian with 
energy $\epsilon^0(g')$. The
difference (including the energy shift by subtracting $\Delta$) yields
$w_{\text{min}}=\epsilon^0(g')-(-\epsilon^0(g))-\Delta=2\epsilon^0(g)$. An analogous explanation
holds for the upper bound.
\subsection{Rate function for the double quench}
\begin{figure}[tb]
    \centering
    \resizebox{\columnwidth}{!}{\includegraphics{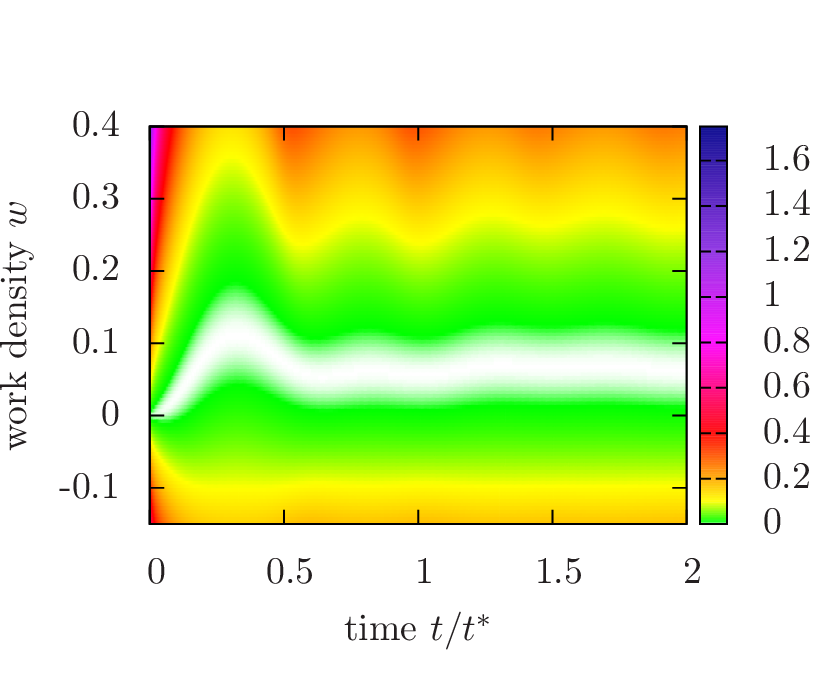}}
    \caption{Rate function $r(w,t)$ for a double quench from $g=0.5$ to $g'=2$ and back to $g$ after
    time $t$ at inverse temperature $\beta=1$ in a color coded form. 
The higher temperature leads to a flatter behavior of the expectation value compared to $\beta=10$, while the rate
functions for fixed times broadens and now almost equally extends to negative $w$. 
In other words, the rate function of the work distribution function becomes more symmetric with
respect to the $w=0$ axis with increasing temperature. This is understood as the result of the
initial canonical distribution where the energy levels become more evenly populated as $T$ grows.}
    \label{fig:time_ratefunction_1}
\end{figure}

In the mentioned work about the DPT Heyl \emph{et al.} computed the rate function of the work
distribution function for a double quench at zero temperature and showed that the DPT is intrinsically included in this
quantity.\cite{heyl2013} In this way, interpreting the Loschmidt echo as the probability density that zero work
is performed, the work distribution function can be regarded as the generalization of the Loschmidt
echo. Their study also used the G\"artner-Ellis theorem to compute the rate function $r(w,t)$ by
transforming the corresponding negative scaled cumulant generating function $k(R,t)$ via a Legendre-Fenchel
transformation as explained earlier. At the critical times $t^*_n=t^*(n+\frac{1}{2})$,
$n=0,1,2\ldots$, the latter quantity showed non-analyticities for quenches across the QCP $g_c=1$
which extended to the rate function if $w=0$, thus rendering $r(w=0,t)$ non-analytic. 
Now, a similar procedure shall investigate whether a DPT can also occur at
non-zero temperatures by analyzing the corresponding negative scaled cumulant generating function
$k(R,t)$. If this function features non-analytic behavior, it will continue to the LF-transformed
version, the rate function. As in the single quench case one therefore studies the integrand in
Eq.\,\eqref{eq:cfowdoublefinal} for possible roots when $u=iR$ with $R\in\mathbb{R}$. However, in
contrast to the $T=0$ case, one immediately perceives that the integrand does not allow for any
zeros, because the denominator consists of a sum of positive numbers which is always $\geq 1$. 
The canonical distribution of initial states represented by the $\cosh(\beta E_k(g))$ terms 
inhibits the existence of roots in any form. In fact, the expression is everywhere analytic as in 
the single quench case for $T>0$. It follows that non-analyticities and therefore dynamical phase 
transitions as in the understanding of Heyl \emph{et al.} cannot occur under any conditions if 
$T>0$ in the transverse field Ising model. \\
\begin{figure}[tb]
    \centering
    \resizebox{\columnwidth}{!}{\includegraphics{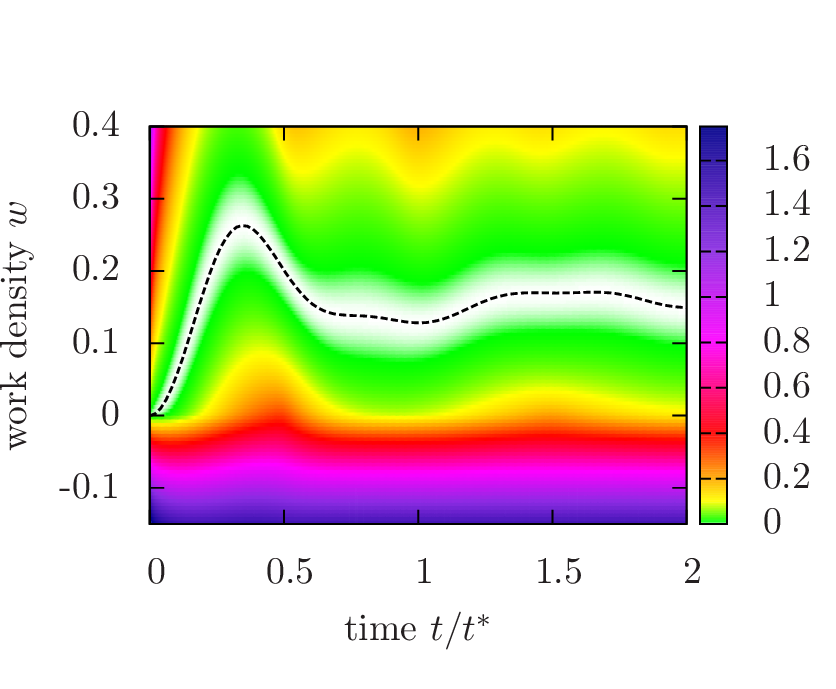}}
    \caption{Rate function $r(w,t)$ for a double quench from $g=0.5$ to $g'=2$ and back to $g$ after
    time $t$ at inverse temperature $\beta=10$ in a color coded form. The dashed line depicts the
expectation value of the performed work, i.\,e. where $r(w,t)=0$. In contrast to the zero
temperature version of this plot the work density plane is extended to negative work densities.
This results from the initial canonical distribution of
states, i.\,e. the inclusion of excited states, such that energy is extracted from the system when the
final state is energetically lower than the initial one.
However, in contrast to positive work densities the ascent of $r(w,t)$ is much steeper for $w<0$
similar to the single quench in Fig.\,\ref{fig:ratefuncsinglezoom}. The red areas around $t/t*=0.5$ and
1.5 are remnants of
the dynamical phase transition at $T=0$ as can be seen in direct comparison and analysing the
constant work density cuts in Fig.\,\ref{fig:horzcuts_beta_10}.}
    \label{fig:time_ratefunction_10_old}
\end{figure}
The asymptotic behavior of $k(R,t)$ is again determined by the ground state energy densities of
the initial and final Hamiltonian which are identical for a double quench, i.\,e., $H(g)$, such that
\begin{align}
    k(R,t)\sim\begin{cases} \,\;\;2\epsilon^0(g)R+C_{\infty} & {\rm for}~R\to\infty\\
        -2\epsilon^0(g)R+C_{-\infty} & {\rm for}~R\to-\infty
    \end{cases}.\label{eq:limitsdouble}
\end{align}
Similarly as before, it follows that $p(w)=0$ everywhere except for $2\epsilon^0(g)\leq w
\leq -2\epsilon^0(g))$. \\
Fig.\,\ref{fig:time_ratefunction_10_old} shows the result for the rate function $r(w,t)$ of a double quench 
protocol across the QCP with identical parameters as in Ref.\,\onlinecite{heyl2013}, but for a low,
non-zero temperature. 
Comparing this result with
its zero temperature correspondence in Fig.\,2 in Ref.\,\onlinecite{heyl2013} one immediately
notices the general qualitative agreement in shape. The expectation value where $r(w,t)=0$ (indicated
by the dashed line) continues to be shifted towards higher work densities at times $t_1^*=0.5$
and $t^*_2=1.5$, albeit the absence of a DPT. 
The new feature is the expected extension to negative work densities which seem to be not as heavily
influenced by the DPT as the positive part and display a steep ascent with decreasing $w$ as in
the single quench case. It is possible that the large slope undermines the visible effects of the DPT. 
Horizontal cuts of the plot for
fixed values $w$ are depicted in Fig.\,\ref{fig:horzcuts_beta_10}. At zero temperature the $w=0$ line 
represents the rate function of the Loschmidt echo and featured the DPT
through kinks at the specified times $t^*_n$. If $T>0$ these kinks cease to exist as expected 
and the rate function $r(w=0,t)$ is smooth.\\
\begin{figure}[tb]
    \centering
    \resizebox{\columnwidth}{!}{\includegraphics{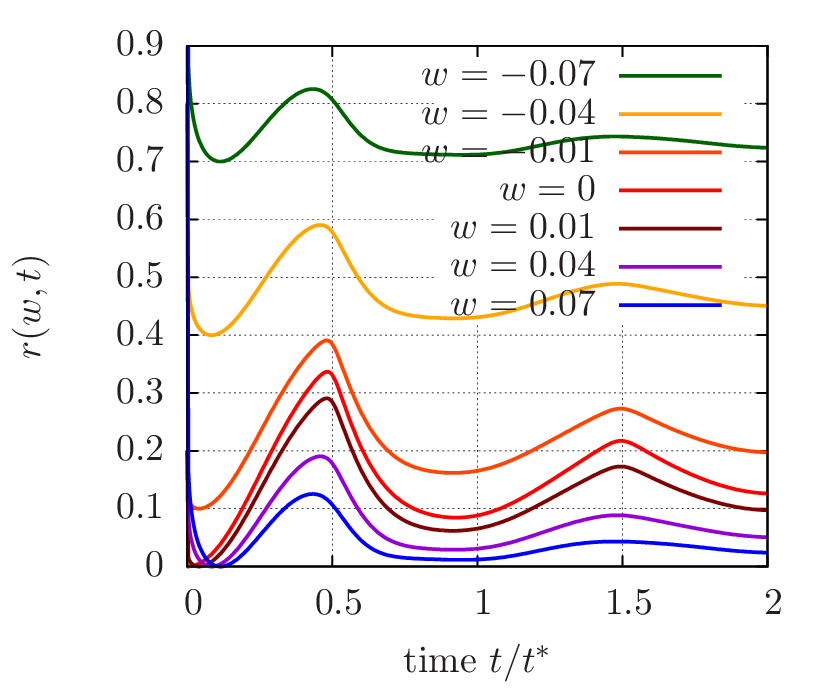}}
    \caption{Various cuts of the rate function plot in Fig.\,\ref{fig:time_ratefunction_10_old} for
fixed values $w$. While the rate function of the Loschmidt echo ($w=0$) at $T=0$ features DPTs
visible as kinks, its non-zero temperature equivalent is perfectly smooth. However, the DPT still
heavily influences the behavior of the rate
function in the form of a shift to larger values of $r(w,t)$.}
    \label{fig:horzcuts_beta_10}
\end{figure}
With increasing temperature the effect of the DPT at $T=0$ diminishes as can be seen in
Fig.\,\ref{fig:time_ratefunction_1}. A higher temperature means that the canonical distribution is
more uniformly distributed such that the number of energy extraction and insertion transitions
equalizes. As a consequence, the rate function becomes more symmetric around the expectation value
which approaches the $w=0$ line. The only visible remnant of the DPT are the slightly shifted rate
functions in the vicinity of $t/t^*\approx 0.25$. In comparison to lower temperatures it appears
that this area has also moved to earlier times. 
\section{\label{sec:tasaki}The Tasaki-Crooks-Jarzynski theorem}
In the last part we study the one-dimensional transverse field Ising model in the context of a
famous fluctuation relation, namely the Tasaki-Crooks-Jarzynski relation.
In general, fluctuation theorems provide additional information about the statistical
properties of the physical quantity of interest and become important whenever a statistical
description is needed for a complete characterization.\cite{campisi2011}
This is, for example, the case when work measurements are considered due to the initial
distribution of states at a certain temperature.
One famous achievement in this context was certainly the \emph{Jarzynski equality}
given by
\begin{align}
    \langle e^{-\beta W}\rangle_{\lambda}=e^{-\beta \Delta F}\label{eq:jarzynski}
\end{align}
in its quantum form for a ramp protocol $\lambda$ where the average is taken with respect to the
initial Gibbs distribution.\cite{jarzynski1997, jarzynskia2008, jarzynski2013} It means that the
mean of exponentiated work values $W$ of many identical experiments equals the exponentiated 
change of the free energy $\Delta F$ of the corresponding equilibrium states. 
The impact of this theorem relies on the fact that
it is also holds for non-equilibrium work measurements. One can therefore learn
about the change of the free energy of the equilibrium states even if the system is not in
equilibrium during the work measurements. A more general theorem including
Eq.\,\eqref{eq:jarzynski} is the \emph{Tasaki-Crooks-Jarzynski relation} 
which relates the work distribution functions of a quench process $\lambda$ and its time-inverted version $\tilde{\lambda}$ to the change of the equilibrium Gibbs free
energy $\Delta F$. It reads
\begin{align}
    \frac{p[w;\lambda]}{p[-w,\tilde{\lambda}]}=e^{N\beta(w-\Delta f)} \label{eq:tasaki}
\end{align}
where $p[w;\lambda]$ denotes the work distribution function of the work density
$w$, $N$ the system size, and $\Delta f=\Delta F/N$ the free energy density.\cite{crooks1999, tasaki2000, kurchan2000}
\subsection{Single quench}
\begin{figure}[tb]
    \centering
    \resizebox{\columnwidth}{!}{\includegraphics{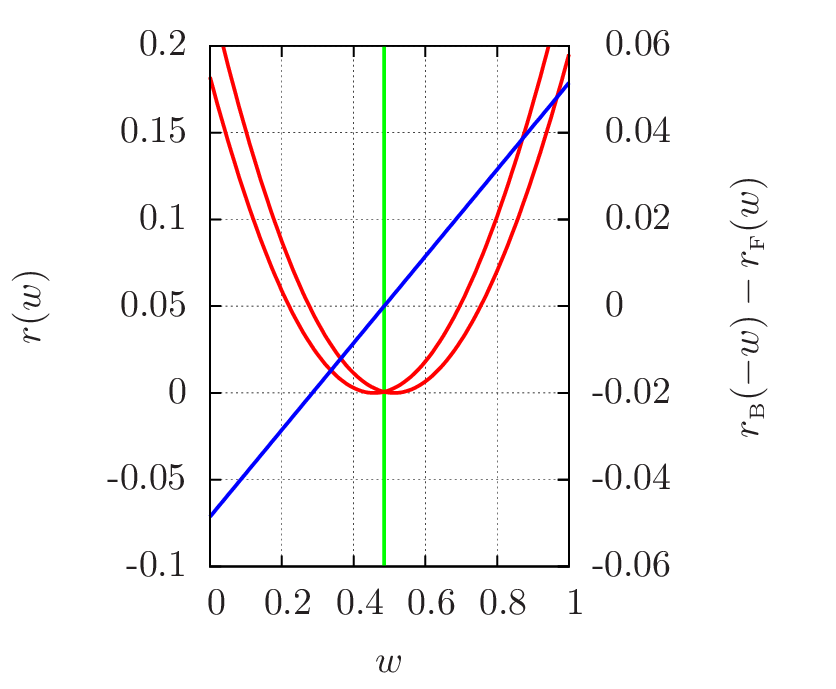}}
    \caption{Rate functions $r(w)$ for a single quench from $g=0.5$ to $g'=2$ (red solid) and its time-reversed 
    counterpart (red dashed) at temperature $\beta=0.1$. The difference of the rate functions shows
linear behavior (blue solid) and allows to directly determine $\Delta f$. The green vertical line
depicts the exact calculation using Eq.\,\eqref{eq:singlecrooksexact}.}
    \label{fig:crookssingle}
\end{figure}

Rewriting
Eq.\,\eqref{eq:tasaki} in terms of its rate functions of the forward (F) and backward (B) quench produces a remarkable result: The Tasaki-Crooks-Jarzynski relation naturally translates into a
simple algebraic equation independent of the system size $N$, namely
\begin{align}
    \frac{1}{N}\ln\left(\frac{p_{\text{F}}(w)}{p_{\text{B}}(-w)}\right)=r_{\text{B}}(-w)-r_{\text{F}}(w)=\beta(w-\Delta f).
\end{align}
This form offers an easy way to measure $\Delta f$ as already shown in experiments.\cite{collin2005,
douarche2005, liphardt2002} 
According to the equation above it is even possible to ``read off''
the value for $\Delta f$ in the plot which is demonstrated in Fig.\,\ref{fig:crookssingle}. The result can be verified
with an exact calculation with the free energy density
\begin{align}
    f(g)=&-\frac{1}{N\beta}\ln Z(g)=-\frac{1}{N\beta}\prod\limits_{0<k<\pi}\left(2+2\cosh\left(\beta
    E_k(g)\right)\right)\nonumber\\
    =&-\frac{1}{2\pi\beta}\int\limits_0^{\pi}\D k\,\ln\left(2+2\cosh\left(\beta
    E_k(g)\right)\right).\label{eq:singlecrooksexact}
\end{align}
Within numerical accuracy both methods show the expected perfect agreement. \\
Beyond numerics the Tasaki-Crooks-Jarzynski relation can be verified exactly using the Fourier transformed version of
Eq.\,\eqref{eq:tasaki}\cite{campisi2011}
\begin{align}
    Z(g)G_{\text{F}}(u)=Z(g')G_{\text{B}}(-u+i\beta). \label{eq:tcjfourier}
\end{align}
On both sides the partition function $Z$ cancels with itsself when plugging in $G(u)$ as in
Eq.\,\eqref{eq:res1} such that only the products over $k$ remain. It follows that the equation above reduces
to an $k$-sector-like form. 
After dividing by 2 on both sides it now reads
\begin{widetext}
\begin{align}
    &1+\cosh\left( (iu+\beta)\epsilon_k(g)
    \right)\cos\left(u\epsilon_k(g')\right)
        -i\,\sinh\left(
    (iu+\beta)\epsilon_k(g) \right)\cos(2\phi_k)\sin\left(u\epsilon_k(g')\right)
    \nonumber\\
    =\;&1+\cosh\left( (i(-u+i\beta)+\beta)\epsilon_k(g')
    \right)\cos\left( (-u+i\beta)\epsilon_k(g)\right)\nonumber\\
        &-i\,\sinh\left(
    (i(-u+i\beta)+\beta)\epsilon_k(g') \right)\cos(-2\phi_k)\sin\left(
    (-u+i\beta)\epsilon_k(g)\right).
\end{align}
\end{widetext}
In the next step one simply uses the trigonometric identities $\cosh(\pm ix)=\cos(x)$ and $\sinh(\pm
ix)=\sin(\pm x)$ with $x\in\mathbb{C}$ to show that the right hand side recovers the left hand side.
Note that the minus sign in the $\cos(-2\phi_k)$ which results from the backward protocol $g'\to g$
is canceled due to the symmetry relation of the $\cos$-function.
\subsection{Double quench}
\begin{figure}[tb]
    \centering
    \resizebox{\columnwidth}{!}{\includegraphics{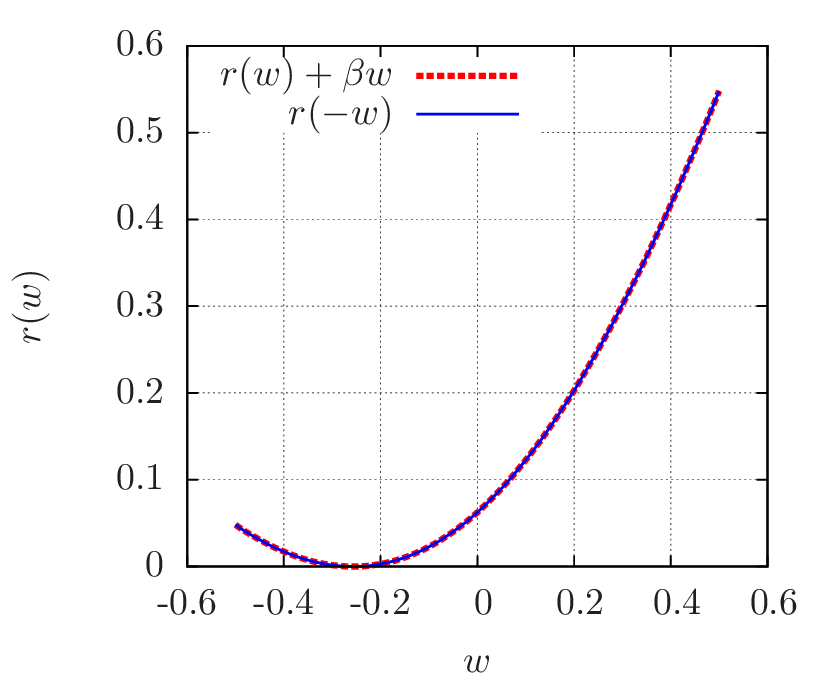}}
    \caption{Rate function $r(-w)$ and $r(w)+\beta w$ for a double quench from $g=0.5$ to $g'=2$ and the
        inverse temperature $\beta = 2$ which are identical. From the
    Tasaki-Crooks-Jarzynski theorem follows that both rate functions are related via the additional
summand $\beta w$, such that $r(w\geq0,t)$ determines $r(-w,t)$ and vice versa.}
    \label{fig:crooksdouble}
\end{figure}

For the double quench Eq.\,\eqref{eq:tasaki} simplifies immediately, because the forward and the
backward quench are identical, i.\,e. the protocol describes a \emph{cyclic process}. Consequently,
$\Delta F=0$ and the equation relates the region $r(w\geq 0)$ to the $r(w<0)$ part:
\begin{align}
    r(w)+\beta w=r(-w). \label{eq:doublecrooks}
\end{align}
It follows that knowing one part of the entire rate function $r(w,t)$ in
Fig.\,\ref{fig:time_ratefunction_10_old} is sufficient, because the other part is determined by it. Especially the
region where $w<0$ and the rate function strongly ascends is difficult to address experimentally.
With the help of Eq.\,\eqref{eq:doublecrooks} it would only be necessary to measure only the part where
the work measurements take on positive values. The numerical verification of the Tasaki-Crooks-Jarzynski
relation is shown in Fig.\,\ref{fig:crooksdouble} where the agreement between the left and right-hand side of
Eq.\,\eqref{eq:doublecrooks} is displayed.
Analogously to the single quench analysis, one can verify the Tasaki-Crooks-Jarzynski theorem
exactly via its Fourier transformed version. Due to the simple symmetric form of the ramp protocol
the partition functions on the left and right side of Eq.\,\eqref{eq:tcjfourier} are identical. 
One can therefore immediately study the final form of the characteristic function of work given in
Eq.\,\eqref{eq:cfowdoublefinal} and check whether it is identical for $u$ and $-u+i\beta$. 
The only part where $u$ appears is the $\cosh$ term and since
\begin{align}
    \cosh\left( (2iu+\beta)\epsilon_k(g) \right)=\cosh\left( (2i(-u+i\beta)+\beta)\epsilon_k(g)
    \right)
\end{align}
the identity is shown.
\section{\label{sec:conclusions}Conclusions}
In this paper we demonstrated analytically that DPTs in the one dimensional transverse field Ising model cannot 
occur at non-zero temperatures due to the initial distribution of states being thermal. To show this
we calculated the characteristic function of work for the simple single and double quench
protocol. Due to the large deviation form of this function it was possible to compute the rate
function of the work distribution function via a Legendre-Fenchel transformation.
We elucidated how the quench parameters define the limits of the rate function in the context of this
transformation. 
Despite the lack of non-analyticities, the work distribution function at $T>0$ 
was shown to be continuously influenced by the DPT at $T=0$. In the quantitative study we
displayed how with increasing temperature the influence diminishes and the rate function changes. 
In the last part we discussed and analytically verified the Tasaki-Crooks-Jarzynski 
theorem for a global single and double quench in the chosen system. Here, the numerical study focused on the 
applicability of the Tasaki-Crooks-Jarzynski relation for different practical purposes.
\begin{acknowledgements}
This work was supported through CRC SFB 1073 (Project B03) of the Deutsche Forschungsgemeinschaft
(DFG). We are grateful to M. Schmitt and D. Fioretto for careful reading of the manuscript and
valuable discussions.

\end{acknowledgements}

\bibliography{lit}

\begin{thebibliography}{31}%
\makeatletter
\providecommand \@ifxundefined [1]{%
 \@ifx{#1\undefined}
}%
\providecommand \@ifnum [1]{%
 \ifnum #1\expandafter \@firstoftwo
 \else \expandafter \@secondoftwo
 \fi
}%
\providecommand \@ifx [1]{%
 \ifx #1\expandafter \@firstoftwo
 \else \expandafter \@secondoftwo
 \fi
}%
\providecommand \natexlab [1]{#1}%
\providecommand \enquote  [1]{``#1''}%
\providecommand \bibnamefont  [1]{#1}%
\providecommand \bibfnamefont [1]{#1}%
\providecommand \citenamefont [1]{#1}%
\providecommand \href@noop [0]{\@secondoftwo}%
\providecommand \href [0]{\begingroup \@sanitize@url \@href}%
\providecommand \@href[1]{\@@startlink{#1}\@@href}%
\providecommand \@@href[1]{\endgroup#1\@@endlink}%
\providecommand \@sanitize@url [0]{\catcode `\\12\catcode `\$12\catcode
  `\&12\catcode `\#12\catcode `\^12\catcode `\_12\catcode `\%12\relax}%
\providecommand \@@startlink[1]{}%
\providecommand \@@endlink[0]{}%
\providecommand \url  [0]{\begingroup\@sanitize@url \@url }%
\providecommand \@url [1]{\endgroup\@href {#1}{\urlprefix }}%
\providecommand \urlprefix  [0]{URL }%
\providecommand \Eprint [0]{\href }%
\providecommand \doibase [0]{http://dx.doi.org/}%
\providecommand \selectlanguage [0]{\@gobble}%
\providecommand \bibinfo  [0]{\@secondoftwo}%
\providecommand \bibfield  [0]{\@secondoftwo}%
\providecommand \translation [1]{[#1]}%
\providecommand \BibitemOpen [0]{}%
\providecommand \bibitemStop [0]{}%
\providecommand \bibitemNoStop [0]{.\EOS\space}%
\providecommand \EOS [0]{\spacefactor3000\relax}%
\providecommand \BibitemShut  [1]{\csname bibitem#1\endcsname}%
\let\auto@bib@innerbib\@empty
\bibitem [{\citenamefont {Polkovnikov}\ \emph {et~al.}(2011)\citenamefont
  {Polkovnikov}, \citenamefont {Sengupta}, \citenamefont {Silva},\ and\
  \citenamefont {Vengalattore}}]{polkovnikov2011}%
  \BibitemOpen
  \bibfield  {author} {\bibinfo {author} {\bibfnamefont {A.}~\bibnamefont
  {Polkovnikov}}, \bibinfo {author} {\bibfnamefont {K.}~\bibnamefont
  {Sengupta}}, \bibinfo {author} {\bibfnamefont {A.}~\bibnamefont {Silva}}, \
  and\ \bibinfo {author} {\bibfnamefont {M.}~\bibnamefont {Vengalattore}},\
  }\href {\doibase 10.1103/RevModPhys.83.863} {\bibfield  {journal} {\bibinfo
  {journal} {Rev. Mod. Phys.}\ }\textbf {\bibinfo {volume} {83}},\ \bibinfo
  {pages} {863} (\bibinfo {year} {2011})}\BibitemShut {NoStop}%
\bibitem [{\citenamefont {Kinoshita}\ \emph {et~al.}(2006)\citenamefont
  {Kinoshita}, \citenamefont {Wenger},\ and\ \citenamefont
  {Weiss}}]{kinoshita2006}%
  \BibitemOpen
  \bibfield  {author} {\bibinfo {author} {\bibfnamefont {T.}~\bibnamefont
  {Kinoshita}}, \bibinfo {author} {\bibfnamefont {T.}~\bibnamefont {Wenger}}, \
  and\ \bibinfo {author} {\bibfnamefont {D.~S.}\ \bibnamefont {Weiss}},\
  }\href@noop {} {\bibfield  {journal} {\bibinfo  {journal} {Nature}\ }\textbf
  {\bibinfo {volume} {440}},\ \bibinfo {pages} {900} (\bibinfo {year}
  {2006})}\BibitemShut {NoStop}%
\bibitem [{\citenamefont {Greiner}\ \emph {et~al.}(2002)\citenamefont
  {Greiner}, \citenamefont {Mandel}, \citenamefont {Esslinger}, \citenamefont
  {H{\"a}nsch},\ and\ \citenamefont {Bloch}}]{greiner2002}%
  \BibitemOpen
  \bibfield  {author} {\bibinfo {author} {\bibfnamefont {M.}~\bibnamefont
  {Greiner}}, \bibinfo {author} {\bibfnamefont {O.}~\bibnamefont {Mandel}},
  \bibinfo {author} {\bibfnamefont {T.}~\bibnamefont {Esslinger}}, \bibinfo
  {author} {\bibfnamefont {T.~W.}\ \bibnamefont {H{\"a}nsch}}, \ and\ \bibinfo
  {author} {\bibfnamefont {I.}~\bibnamefont {Bloch}},\ }\href@noop {}
  {\bibfield  {journal} {\bibinfo  {journal} {Nature}\ }\textbf {\bibinfo
  {volume} {415}},\ \bibinfo {pages} {39} (\bibinfo {year} {2002})}\BibitemShut
  {NoStop}%
\bibitem [{\citenamefont {Talkner}\ \emph {et~al.}(2007)\citenamefont
  {Talkner}, \citenamefont {Lutz},\ and\ \citenamefont
  {H{\"a}nggi}}]{talkner2007work}%
  \BibitemOpen
  \bibfield  {author} {\bibinfo {author} {\bibfnamefont {P.}~\bibnamefont
  {Talkner}}, \bibinfo {author} {\bibfnamefont {E.}~\bibnamefont {Lutz}}, \
  and\ \bibinfo {author} {\bibfnamefont {P.}~\bibnamefont {H{\"a}nggi}},\
  }\href {\doibase 10.1103/PhysRevE.75.050102} {\bibfield  {journal} {\bibinfo
  {journal} {Phys. Rev. E}\ }\textbf {\bibinfo {volume} {75}},\ \bibinfo
  {pages} {050102} (\bibinfo {year} {2007})}\BibitemShut {NoStop}%
\bibitem [{\citenamefont {Campisi}\ \emph {et~al.}(2011)\citenamefont
  {Campisi}, \citenamefont {H{\"a}nggi},\ and\ \citenamefont
  {Talkner}}]{campisi2011}%
  \BibitemOpen
  \bibfield  {author} {\bibinfo {author} {\bibfnamefont {M.}~\bibnamefont
  {Campisi}}, \bibinfo {author} {\bibfnamefont {P.}~\bibnamefont {H{\"a}nggi}},
  \ and\ \bibinfo {author} {\bibfnamefont {P.}~\bibnamefont {Talkner}},\ }\href
  {\doibase 10.1103/RevModPhys.83.771} {\bibfield  {journal} {\bibinfo
  {journal} {Rev. Mod. Phys.}\ }\textbf {\bibinfo {volume} {83}},\ \bibinfo
  {pages} {771} (\bibinfo {year} {2011})}\BibitemShut {NoStop}%
\bibitem [{\citenamefont {Liphardt}\ \emph {et~al.}(2002)\citenamefont
  {Liphardt}, \citenamefont {Dumont}, \citenamefont {Smith}, \citenamefont
  {Tinoco},\ and\ \citenamefont {Bustamante}}]{liphardt2002}%
  \BibitemOpen
  \bibfield  {author} {\bibinfo {author} {\bibfnamefont {J.}~\bibnamefont
  {Liphardt}}, \bibinfo {author} {\bibfnamefont {S.}~\bibnamefont {Dumont}},
  \bibinfo {author} {\bibfnamefont {S.~B.}\ \bibnamefont {Smith}}, \bibinfo
  {author} {\bibfnamefont {I.}~\bibnamefont {Tinoco}}, \ and\ \bibinfo {author}
  {\bibfnamefont {C.}~\bibnamefont {Bustamante}},\ }\href@noop {} {\bibfield
  {journal} {\bibinfo  {journal} {Science}\ }\textbf {\bibinfo {volume}
  {296}},\ \bibinfo {pages} {1832} (\bibinfo {year} {2002})}\BibitemShut
  {NoStop}%
\bibitem [{\citenamefont {Collin}\ \emph {et~al.}(2005)\citenamefont {Collin},
  \citenamefont {Ritort}, \citenamefont {Jarzynski}, \citenamefont {Smith},
  \citenamefont {Tinoco},\ and\ \citenamefont {Bustamante}}]{collin2005}%
  \BibitemOpen
  \bibfield  {author} {\bibinfo {author} {\bibfnamefont {D.}~\bibnamefont
  {Collin}}, \bibinfo {author} {\bibfnamefont {F.}~\bibnamefont {Ritort}},
  \bibinfo {author} {\bibfnamefont {C.}~\bibnamefont {Jarzynski}}, \bibinfo
  {author} {\bibfnamefont {S.}~\bibnamefont {Smith}}, \bibinfo {author}
  {\bibfnamefont {I.}~\bibnamefont {Tinoco}}, \ and\ \bibinfo {author}
  {\bibfnamefont {C.}~\bibnamefont {Bustamante}},\ }\href@noop {} {\bibfield
  {journal} {\bibinfo  {journal} {Nature}\ }\textbf {\bibinfo {volume} {437}},\
  \bibinfo {pages} {231} (\bibinfo {year} {2005})}\BibitemShut {NoStop}%
\bibitem [{\citenamefont {Douarche}\ \emph {et~al.}(2005)\citenamefont
  {Douarche}, \citenamefont {Ciliberto}, \citenamefont {Petrosyan},\ and\
  \citenamefont {Rabbiosi}}]{douarche2005}%
  \BibitemOpen
  \bibfield  {author} {\bibinfo {author} {\bibfnamefont {F.}~\bibnamefont
  {Douarche}}, \bibinfo {author} {\bibfnamefont {S.}~\bibnamefont {Ciliberto}},
  \bibinfo {author} {\bibfnamefont {A.}~\bibnamefont {Petrosyan}}, \ and\
  \bibinfo {author} {\bibfnamefont {I.}~\bibnamefont {Rabbiosi}},\ }\href@noop
  {} {\bibfield  {journal} {\bibinfo  {journal} {Europhys. Lett.}\ }\textbf
  {\bibinfo {volume} {70}},\ \bibinfo {pages} {593} (\bibinfo {year}
  {2005})}\BibitemShut {NoStop}%
\bibitem [{\citenamefont {An}\ \emph {et~al.}(2015)\citenamefont {An},
  \citenamefont {Zhang}, \citenamefont {Um}, \citenamefont {Lv}, \citenamefont
  {Lu}, \citenamefont {Zhang}, \citenamefont {Yin}, \citenamefont {Quan},\ and\
  \citenamefont {Kim}}]{an2015}%
  \BibitemOpen
  \bibfield  {author} {\bibinfo {author} {\bibfnamefont {S.}~\bibnamefont
  {An}}, \bibinfo {author} {\bibfnamefont {J.-N.}\ \bibnamefont {Zhang}},
  \bibinfo {author} {\bibfnamefont {M.}~\bibnamefont {Um}}, \bibinfo {author}
  {\bibfnamefont {D.}~\bibnamefont {Lv}}, \bibinfo {author} {\bibfnamefont
  {Y.}~\bibnamefont {Lu}}, \bibinfo {author} {\bibfnamefont {J.}~\bibnamefont
  {Zhang}}, \bibinfo {author} {\bibfnamefont {Z.-Q.}\ \bibnamefont {Yin}},
  \bibinfo {author} {\bibfnamefont {H.~T.}\ \bibnamefont {Quan}}, \ and\
  \bibinfo {author} {\bibfnamefont {K.}~\bibnamefont {Kim}},\ }\href {\doibase
  10.1038/nphys3197} {\bibfield  {journal} {\bibinfo  {journal} {Nature
  Physics}\ }\textbf {\bibinfo {volume} {11}},\ \bibinfo {pages} {193}
  (\bibinfo {year} {2015})}\BibitemShut {NoStop}%
\bibitem [{\citenamefont {Heyl}\ \emph {et~al.}(2013)\citenamefont {Heyl},
  \citenamefont {Polkovnikov},\ and\ \citenamefont {Kehrein}}]{heyl2013}%
  \BibitemOpen
  \bibfield  {author} {\bibinfo {author} {\bibfnamefont {M.}~\bibnamefont
  {Heyl}}, \bibinfo {author} {\bibfnamefont {A.}~\bibnamefont {Polkovnikov}}, \
  and\ \bibinfo {author} {\bibfnamefont {S.}~\bibnamefont {Kehrein}},\ }\href
  {\doibase 10.1103/PhysRevLett.110.135704} {\bibfield  {journal} {\bibinfo
  {journal} {Phys. Rev. Lett.}\ }\textbf {\bibinfo {volume} {110}},\ \bibinfo
  {pages} {135704} (\bibinfo {year} {2013})}\BibitemShut {NoStop}%
\bibitem [{\citenamefont {Silva}(2008)}]{silva2008}%
  \BibitemOpen
  \bibfield  {author} {\bibinfo {author} {\bibfnamefont {A.}~\bibnamefont
  {Silva}},\ }\href {\doibase 10.1103/PhysRevLett.101.120603} {\bibfield
  {journal} {\bibinfo  {journal} {Phys. Rev. Lett.}\ }\textbf {\bibinfo
  {volume} {101}},\ \bibinfo {pages} {120603} (\bibinfo {year}
  {2008})}\BibitemShut {NoStop}%
\bibitem [{\citenamefont {Karrasch}\ and\ \citenamefont
  {Schuricht}(2013)}]{karrasch2013}%
  \BibitemOpen
  \bibfield  {author} {\bibinfo {author} {\bibfnamefont {C.}~\bibnamefont
  {Karrasch}}\ and\ \bibinfo {author} {\bibfnamefont {D.}~\bibnamefont
  {Schuricht}},\ }\href {\doibase 10.1103/PhysRevB.87.195104} {\bibfield
  {journal} {\bibinfo  {journal} {Phys. Rev. B}\ }\textbf {\bibinfo {volume}
  {87}},\ \bibinfo {pages} {195104} (\bibinfo {year} {2013})}\BibitemShut
  {NoStop}%
\bibitem [{\citenamefont {Kriel}\ \emph {et~al.}(2014)\citenamefont {Kriel},
  \citenamefont {Karrasch},\ and\ \citenamefont {Kehrein}}]{kriel2014}%
  \BibitemOpen
  \bibfield  {author} {\bibinfo {author} {\bibfnamefont {J.~N.}\ \bibnamefont
  {Kriel}}, \bibinfo {author} {\bibfnamefont {C.}~\bibnamefont {Karrasch}}, \
  and\ \bibinfo {author} {\bibfnamefont {S.}~\bibnamefont {Kehrein}},\ }\href
  {\doibase 10.1103/PhysRevB.90.125106} {\bibfield  {journal} {\bibinfo
  {journal} {Phys. Rev. B}\ }\textbf {\bibinfo {volume} {90}},\ \bibinfo
  {pages} {125106} (\bibinfo {year} {2014})}\BibitemShut {NoStop}%
\bibitem [{\citenamefont {Vajna}\ and\ \citenamefont
  {D\'ora}(2014)}]{vajna2014}%
  \BibitemOpen
  \bibfield  {author} {\bibinfo {author} {\bibfnamefont {S.}~\bibnamefont
  {Vajna}}\ and\ \bibinfo {author} {\bibfnamefont {B.}~\bibnamefont {D\'ora}},\
  }\href {\doibase 10.1103/PhysRevB.89.161105} {\bibfield  {journal} {\bibinfo
  {journal} {Phys. Rev. B}\ }\textbf {\bibinfo {volume} {89}},\ \bibinfo
  {pages} {161105} (\bibinfo {year} {2014})}\BibitemShut {NoStop}%
\bibitem [{\citenamefont {Andraschko}\ and\ \citenamefont
  {Sirker}(2014)}]{andraschko2014}%
  \BibitemOpen
  \bibfield  {author} {\bibinfo {author} {\bibfnamefont {F.}~\bibnamefont
  {Andraschko}}\ and\ \bibinfo {author} {\bibfnamefont {J.}~\bibnamefont
  {Sirker}},\ }\href {\doibase 10.1103/PhysRevB.89.125120} {\bibfield
  {journal} {\bibinfo  {journal} {Phys. Rev. B}\ }\textbf {\bibinfo {volume}
  {89}},\ \bibinfo {pages} {125120} (\bibinfo {year} {2014})}\BibitemShut
  {NoStop}%
\bibitem [{\citenamefont {Canovi}\ \emph {et~al.}(2014)\citenamefont {Canovi},
  \citenamefont {Werner},\ and\ \citenamefont {Eckstein}}]{canovi2014}%
  \BibitemOpen
  \bibfield  {author} {\bibinfo {author} {\bibfnamefont {E.}~\bibnamefont
  {Canovi}}, \bibinfo {author} {\bibfnamefont {P.}~\bibnamefont {Werner}}, \
  and\ \bibinfo {author} {\bibfnamefont {M.}~\bibnamefont {Eckstein}},\ }\href
  {\doibase 10.1103/PhysRevLett.113.265702} {\bibfield  {journal} {\bibinfo
  {journal} {Phys. Rev. Lett.}\ }\textbf {\bibinfo {volume} {113}},\ \bibinfo
  {pages} {265702} (\bibinfo {year} {2014})}\BibitemShut {NoStop}%
\bibitem [{\citenamefont {Fagotti}(2013)}]{fagotti2013}%
  \BibitemOpen
  \bibfield  {author} {\bibinfo {author} {\bibfnamefont {M.}~\bibnamefont
  {Fagotti}},\ }\href@noop {} {\bibfield  {journal} {\bibinfo  {journal} {ArXiv
  e-prints}\ } (\bibinfo {year} {2013})},\ \Eprint
  {http://arxiv.org/abs/1308.0277} {arXiv:1308.0277 [cond-mat.stat-mech]}
  \BibitemShut {NoStop}%
\bibitem [{\citenamefont {Heyl}(2014)}]{heyl2014}%
  \BibitemOpen
  \bibfield  {author} {\bibinfo {author} {\bibfnamefont {M.}~\bibnamefont
  {Heyl}},\ }\href {\doibase 10.1103/PhysRevLett.113.205701} {\bibfield
  {journal} {\bibinfo  {journal} {Phys. Rev. Lett.}\ }\textbf {\bibinfo
  {volume} {113}},\ \bibinfo {pages} {205701} (\bibinfo {year}
  {2014})}\BibitemShut {NoStop}%
\bibitem [{\citenamefont {Heyl}(2015)}]{heyl2015universality}%
  \BibitemOpen
  \bibfield  {author} {\bibinfo {author} {\bibfnamefont {M.}~\bibnamefont
  {Heyl}},\ }\href {\doibase 10.1103/PhysRevLett.115.140602} {\bibfield
  {journal} {\bibinfo  {journal} {Phys. Rev. Lett.}\ }\textbf {\bibinfo
  {volume} {115}},\ \bibinfo {pages} {140602} (\bibinfo {year}
  {2015})}\BibitemShut {NoStop}%
\bibitem [{\citenamefont {Andrieux}\ and\ \citenamefont
  {Gaspard}(2008)}]{andrieux2008}%
  \BibitemOpen
  \bibfield  {author} {\bibinfo {author} {\bibfnamefont {D.}~\bibnamefont
  {Andrieux}}\ and\ \bibinfo {author} {\bibfnamefont {P.}~\bibnamefont
  {Gaspard}},\ }\href@noop {} {\bibfield  {journal} {\bibinfo  {journal} {Phys.
  Rev. Lett.}\ }\textbf {\bibinfo {volume} {100}},\ \bibinfo {pages} {230404}
  (\bibinfo {year} {2008})}\BibitemShut {NoStop}%
\bibitem [{\citenamefont {Smacchia}\ and\ \citenamefont
  {Silva}(2013)}]{smacchia2013}%
  \BibitemOpen
  \bibfield  {author} {\bibinfo {author} {\bibfnamefont {P.}~\bibnamefont
  {Smacchia}}\ and\ \bibinfo {author} {\bibfnamefont {A.}~\bibnamefont
  {Silva}},\ }\href {\doibase 10.1103/PhysRevE.88.042109} {\bibfield  {journal}
  {\bibinfo  {journal} {Phys. Rev. E}\ }\textbf {\bibinfo {volume} {88}},\
  \bibinfo {pages} {042109} (\bibinfo {year} {2013})}\BibitemShut {NoStop}%
\bibitem [{\citenamefont {Sachdev}(2011)}]{sachdev2011}%
  \BibitemOpen
  \bibfield  {author} {\bibinfo {author} {\bibfnamefont {S.}~\bibnamefont
  {Sachdev}},\ }\href@noop {} {\emph {\bibinfo {title} {Quantum Phase
  Transitions}}},\ \bibinfo {edition} {2nd}\ ed.\ (\bibinfo  {publisher}
  {Cambridge Univ. Press},\ \bibinfo {address} {Cambridge},\ \bibinfo {year}
  {2011})\BibitemShut {NoStop}%
\bibitem [{\citenamefont {Lieb}\ \emph {et~al.}(1961)\citenamefont {Lieb},
  \citenamefont {Schultz},\ and\ \citenamefont {Mattis}}]{lieb1961}%
  \BibitemOpen
  \bibfield  {author} {\bibinfo {author} {\bibfnamefont {E.}~\bibnamefont
  {Lieb}}, \bibinfo {author} {\bibfnamefont {T.}~\bibnamefont {Schultz}}, \
  and\ \bibinfo {author} {\bibfnamefont {D.}~\bibnamefont {Mattis}},\
  }\href@noop {} {\bibfield  {journal} {\bibinfo  {journal} {Ann. Phys.(NY)}\
  }\textbf {\bibinfo {volume} {16}},\ \bibinfo {pages} {407} (\bibinfo {year}
  {1961})}\BibitemShut {NoStop}%
\bibitem [{\citenamefont {Gambassi}\ and\ \citenamefont
  {Silva}(2012)}]{gambassi2012}%
  \BibitemOpen
  \bibfield  {author} {\bibinfo {author} {\bibfnamefont {A.}~\bibnamefont
  {Gambassi}}\ and\ \bibinfo {author} {\bibfnamefont {A.}~\bibnamefont
  {Silva}},\ }\href {\doibase 10.1103/PhysRevLett.109.250602} {\bibfield
  {journal} {\bibinfo  {journal} {Phys. Rev. Lett.}\ }\textbf {\bibinfo
  {volume} {109}},\ \bibinfo {pages} {250602} (\bibinfo {year}
  {2012})}\BibitemShut {NoStop}%
\bibitem [{\citenamefont {Touchette}(2009)}]{touchette2009}%
  \BibitemOpen
  \bibfield  {author} {\bibinfo {author} {\bibfnamefont {H.}~\bibnamefont
  {Touchette}},\ }\href {\doibase
  http://dx.doi.org/10.1016/j.physrep.2009.05.002} {\bibfield  {journal}
  {\bibinfo  {journal} {Phys Rep.}\ }\textbf {\bibinfo {volume} {478}},\
  \bibinfo {pages} {1 } (\bibinfo {year} {2009})}\BibitemShut {NoStop}%
\bibitem [{\citenamefont {Jarzynski}(1997)}]{jarzynski1997}%
  \BibitemOpen
  \bibfield  {author} {\bibinfo {author} {\bibfnamefont {C.}~\bibnamefont
  {Jarzynski}},\ }\href@noop {} {\bibfield  {journal} {\bibinfo  {journal}
  {Phys. Rev. Lett.}\ }\textbf {\bibinfo {volume} {78}},\ \bibinfo {pages}
  {2690} (\bibinfo {year} {1997})}\BibitemShut {NoStop}%
\bibitem [{\citenamefont {Jarzynski}(2008)}]{jarzynskia2008}%
  \BibitemOpen
  \bibfield  {author} {\bibinfo {author} {\bibfnamefont {C.}~\bibnamefont
  {Jarzynski}},\ }\href@noop {} {\bibfield  {journal} {\bibinfo  {journal}
  {Euro. Phys. J. B}\ }\textbf {\bibinfo {volume} {64}},\ \bibinfo {pages}
  {331} (\bibinfo {year} {2008})}\BibitemShut {NoStop}%
\bibitem [{\citenamefont {Jarzynski}(2013)}]{jarzynski2013}%
  \BibitemOpen
  \bibfield  {author} {\bibinfo {author} {\bibfnamefont {C.}~\bibnamefont
  {Jarzynski}},\ }in\ \href {\doibase 10.1007/978-3-0348-0359-5_4} {\emph
  {\bibinfo {booktitle} {Time}}},\ \bibinfo {series} {Progress in Mathematical
  Physics}, Vol.~\bibinfo {volume} {63},\ \bibinfo {editor} {edited by\
  \bibinfo {editor} {\bibfnamefont {B.}~\bibnamefont {Duplantier}}}\ (\bibinfo
  {publisher} {Springer Basel},\ \bibinfo {year} {2013})\ pp.\ \bibinfo {pages}
  {145--172}\BibitemShut {NoStop}%
\bibitem [{\citenamefont {Crooks}(1999)}]{crooks1999}%
  \BibitemOpen
  \bibfield  {author} {\bibinfo {author} {\bibfnamefont {G.~E.}\ \bibnamefont
  {Crooks}},\ }\href@noop {} {\bibfield  {journal} {\bibinfo  {journal} {Phys.
  Rev. E}\ }\textbf {\bibinfo {volume} {60}},\ \bibinfo {pages} {2721}
  (\bibinfo {year} {1999})}\BibitemShut {NoStop}%
\bibitem [{\citenamefont {Tasaki}(2000)}]{tasaki2000}%
  \BibitemOpen
  \bibfield  {author} {\bibinfo {author} {\bibfnamefont {H.}~\bibnamefont
  {Tasaki}},\ }\href@noop {} {\bibfield  {journal} {\bibinfo  {journal} {arXiv
  preprint cond-mat/0009244}\ } (\bibinfo {year} {2000})}\BibitemShut {NoStop}%
\bibitem [{\citenamefont {Kurchan}(2000)}]{kurchan2000}%
  \BibitemOpen
  \bibfield  {author} {\bibinfo {author} {\bibfnamefont {J.}~\bibnamefont
  {Kurchan}},\ }\href@noop {} {\bibfield  {journal} {\bibinfo  {journal} {arXiv
  preprint cond-mat/0007360}\ } (\bibinfo {year} {2000})}\BibitemShut {NoStop}%
\end{thebibliography}%

\end{document}